\documentclass[twocolumn,english]{revtex4}
\usepackage[T1]{fontenc}
\usepackage[latin9]{inputenc}
\usepackage{babel}

\usepackage{graphicx}
\usepackage{amstext}
\usepackage{amssymb}
\usepackage[unicode=true,pdfusetitle,
 bookmarks=true,bookmarksnumbered=false,bookmarksopen=false,
 breaklinks=false,pdfborder={0 0 1},backref=false,colorlinks=false]
 {hyperref}

\begin{document}

\title{Spatial Coherence of a polariton condensate in 1D acoustic
lattice}

\author{O. Tsyplyatyev}
\affiliation{Department of Physics and Astronomy, University of Sheffield, Sheffield
S3 7RH, UK}

\author{D. M. Whittaker}
\affiliation{Department of Physics and Astronomy, University of Sheffield, Sheffield
S3 7RH, UK}

\begin{abstract}
Several mechanisms are discussed which could determine the spatial
coherence of a polariton condensate confined to a one dimensional
wire.  The mechanisms considered are polariton-polariton interactions,
disorder scattering and non-equilibrium occupation of finite momentum
modes. For each case, the shape of the resulting spatial coherence
function $g^{(1)}\left(x\right)$ is analysed. The results are
compared with the experimental data on a polariton condensate in an
acoustic lattice from {[}E. A. Cerda-Mendez \emph{et al},
Phys. Rev. Lett. \textbf{105}, 116402 (2010){]}. It is concluded that
the shape of $g^{(1)}\left(x\right)$ can only be explained by
non-equilibrium effects, and that $\sim 10$ modes are occupied in the
experimental system. 
\end{abstract}

\date{\today}

\maketitle

\section{Introduction}

The fabrication of planar (2D) microcavities has provided experimental
access to the strong coupling regime between light and matter in a
spatially extended system \cite{Weisbuch}. The excitations of the
coupled modes are polaritons, quasi-particles which are bosons with a
very small effective mass, typically $10^{-4}$ times that of an
electron. This makes it possible to study quantum effects in a solid
state system at a relatively high temperatures
\cite{SkolnickFischerWhittaker}.  The short life time of the
polaritons, $\sim5$ ps \cite{Balili}, means that the system has to be constantly pumped to observe Bose-Einstein condensation (BEC). It can be achieved either resonantly \cite{Stevenson} or
non-resonantly \cite{Kasprzak} producing condensates with long coherence times \cite{Love} and long range
spatial coherence \cite{Balili,Kasprzak,Wertz}. 
These have been shown to
exhibit interesting nonlinear and many-body phenomena, such as vortex
formation \cite{Lagoudakis08,Krizhanovskiivortex}, solitons \cite{AmoSoliton}
and superfluid-like flow \cite{AmoSuperfluid}.

In recent experiments \cite{Wertz,Krizhanovskii}, microcavity
polaritons have been confined in wire like geometries, to study the
peculiar interaction effects of bosons in a 1D system as well as a
step to the realisation of polariton circuits \cite{Liew}.  The extra
confinement was provided in two different ways: by etching of the
whole planar structure to form a wire \cite{Wertz} and by excitation
of surface acoustic waves (SAW) to form a dynamic 1D
lattice \cite{Krizhanovskii}. It is the spatial coherence measurements
in the latter setup that are the subject of the present paper.  In the
experiment, the same sample area was studied with and without the
SAW, so it was possible to measure how the coherence length changed
when the 1D confinement was introduced. Without confinement, the
entire 2D emission spot was coherent, that is, the coherence length
was the same as the spot size \cite{Krizhanovskii,Deng2007}. With confinement it was observed that
the coherence length in the direction perpendicular to the SAW
wavefronts was reduced to the SAW wavelength, as the condensates
confined in the individual minima became decoupled. However, there was
also a significant reduction in the coherence length measured
parallel to the wavefronts \cite{Krizhanovskii}, along the `wires', which is less easily
explained.

In this paper we analyse possible mechanisms for the reduction of the
spatial coherence along the 1D wires. We consider the effect \cite{polariton_phonon}  of
polariton-polariton interactions, disorder, and finite momentum modes
occupation in a harmonic trap on the first order correlation function
$g^{(1)}\left(x\right)$. The theoretical predictions are compared with
the experimentally measured $g^{(1)}\left(x\right)$ \cite{exp_data}
which has a Gaussian shape up to the noise floor. This shape can only
be reproduced by the non-equilibrium occupation of a set of the finite
momentum modes. Performing a more quantitative fit, using the
occupation numbers as fitting parameters, we find that the number
of the excited modes in the experiment \cite{Krizhanovskii} is
$\sim10$.

The rest of the paper is organised as follows. In Section 2 we formulate
the model of the 1D polariton system with different types of interactions.
In Section 3 we calculate the spatial dependence of the coherence
function $g^{(1)}\left(x\right)$ when finite momentum modes are occupied
in a harmonic trap. Section 4 contains discussion of the Anderson
localisation due to disorder in a 1D system. In Section 5 we discuss
the effect of the short range polariton-polarion interaction on $g^{(1)}\left(x\right)$.
Section 6 contains the fitting of the experimental data with the
theoretical predictions from the previous sections.

\section{Model}

We consider one dimensional polaritons formed by the equal mixture of
exciton and photon at the zero momentum following the experimental
setup of \cite{Krizhanovskii}.

The motion of photons in $z-$direction is restricted  by a planar
microcavity which is formed by a pair of Bragg mirrors giving a small mass to the 2D photons  $C_{\mathbf{k}}$.
Excitons are bound electron-hole pairs, they are bosons $X_{\mathbf{k}}$,
which are confined to the same 2D plane by a quantum well, see the
scheme in Fig. 1. The eigenmodes of a strongly interacting  $X_{\mathbf{k}}$
and $C_{\mathbf{k}}$ are two hybrid particles, Lower and Upper
polarions (LP and UP). We
will consider only the LP branch as bosons $a_{\mathbf{k}}$ with
a parabolic dispersion and mass $m_{LP}$ assuming that the photons
and the excitons are at resonance at $\mathbf{k}=0$, e. g. a LP around
the zero momentum consists of equal mixture of the exciton and the
photon $a_{\mathbf{k}}=\left(X_{\mathbf{k}}+C_{\mathbf{k}}\right)/\sqrt{2}$.

Bosonic nature of these polaritons allows a macroscopic occupation
of the single zero momentum state that leads to a long-range phase
coherence in the system. Access to this regime is not straightforward
as the short lived polaritons do not have time to relax at low momenta
due to a slow emission of the low energy phonons forming a relaxation
bottleneck and forbidding formation of the truly equilibrium condensate. This difficulty can be overcome in a nonequilibrium regime under a strong pumping when the $\mathbf{k}=0$ state is populated directly or indirectly by employing an optical parametric oscillator (OPO).  

In the OPO scheme the system is excited by the pump at a finite momentum $\mathbf{k}_{p}$
and energy $E_{p}$. Due to the polariton-polariton
interactions the $\mathbf{k}_p$ excitations can scatter fast to a
low momentum. Above a threshold density of the directly pumped polaritons the interaction becomes strong enough to reach the zero momentum state $\mathbf{k}_{s}=0$, signal, which also forms a significant population in the idler state $\mathbf{k}_{i}$ due to the conservation of the
energy $E_{i}+E_{s}=E_{p}$ and the momentum $\mathbf{k}_{i}+\mathbf{k}_{s}=\mathbf{k}_{p}$
by the scattering process. Above the threshold the systems locks only
to these three states \cite{Stevenson} with a significant population
of the signal state at $\mathbf{k}_{s}=0$. This approach is more
advantageous than the direct pumping as in the OPO regime the pump
at a finite $\mathbf{k}_{p}$ does not obstruct the emission from
the condensate at $\mathbf{k}_{s}=0$. 

The 2D polaritons are constrained further to a set of 1D wires by an
in-plane surface acoustic wave propagating along $y$-direction, see
the scheme in Fig. 1. The $y$-component of the momentum belongs
to the lowest energy subband leaving the only motion  along $x$-axis
unconstrained.  The polaritons are detected though the photonic part which constantly escapes the system forming a spatial distribution of the electric field outside of the cavity that contains information about the polaritons inside the structure, 

\begin{equation}
E\left(x\right)=\sum_{k}\sqrt{\frac{\epsilon_{k}}{L}}\left(\varphi_{k}^{*}\left(x\right)C_{k}^{\dagger}-\varphi_{k}\left(x\right)C_{k}\right),\end{equation}
and can be registered by a detector. Here $\varphi_{k}\left(x\right)$
are the eigenmodes of the external potential in $x$-direction, e.
g. $\varphi_{k}\left(x\right)=e^{ikx}$ if the there is no extra potential
along $x$-axis, $\epsilon_{k}$ is the dispersion of the electro-magnetic
modes, and $L$ is the length of the wire.

\begin{figure}[t]
\centering\includegraphics[width=1\columnwidth]{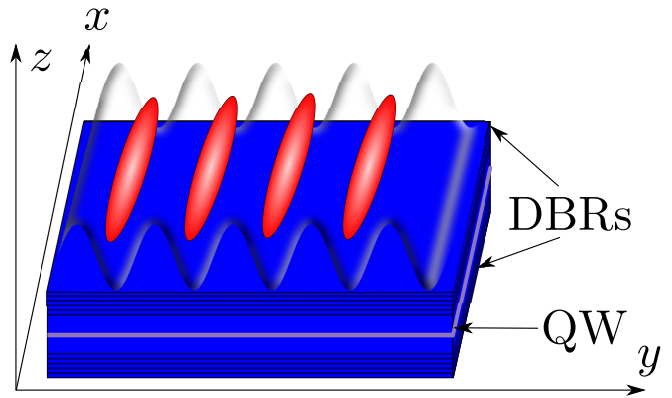}
\caption{\label{fig:experimental_setup} Scheme of the microcavity (DBRs) with a quantum
well inside (QW), and the surface acoustic wave (red elongated ellipses
are the polariton 1D wires).}
\end{figure}

We consider three possible mechanisms that can affect spatial coherence of free polaritons in the signal, $a_{k}^{\dagger}$
around $k_{s}=0$. First is a confining potential originating from
a finite size of the excitation spot. We assume that the single particle
potential is harmonic,\begin{equation}
U_{\textrm{harmonic}}=\frac{m_{LP}\omega^{2}x^{2}}{2},\label{eq:trap_def}\end{equation}
where the frequency $\omega$ defines the system size as $L=\sqrt{2\hbar/\left(m_{LP}\omega\right)}$.
Second is a disorder potential that has a finite correlation length
$l_{0}$, \begin{equation}
\overline{U_{\textrm{disorder}}\left(x\right)U_{\textrm{disorder}}\left(x'\right)}=F\left(\frac{x-x'}{l_{0}}\right)\label{eq:disorder_def}\end{equation}
where $F\left(y\right)$ is a function that vanished when $\left|x-x'\right|\gg l$,
i. e. $F\left(0\right)=U_{0}$ and $F\left(y\gg1\right)=0$. And the
third mechanism is the pair polariton-polariton interaction. It has a short
range interaction potential as the polaritons have zero electric charge
with the interaction strength $V_{0}$.
 
We neglect the interaction of polariton with phonons mechanism for decoherence as it is  ineffective at low momenta. The polariton-phonon scattering time is $\sim 100$  ps \cite{Tassone} which is much longer than the polariton life time $\sim 5$ ps. The latter is due to detuning chosen in the experiment \cite{Krizhanovskii} where the polaritons at $\mathbf{k}\approx 0$ consist of the equal mixture of the exciton and the photon.

To analyse the spatial coherence that is observed via the photonic part
of the LP we consider the first order correlation function, \begin{equation}
g^{(1)}\left(x\right)=\frac{\left\langle \hat{N}\left[E^{\dagger}\left(x\right)E\left(-x\right)\right]\right\rangle }{\left\langle \hat{N}\left[E^{\dagger}\left(0\right)E\left(0\right)\right]\right\rangle },\label{eq:g1_def}\end{equation}
where $\hat{N}\left[\dots\right]$ is the normal ordering operator
that eliminates the vacuum fluctuations of the quantised electro-magnetic
field and  $\left\langle \dots\right\rangle $
is the average with respect to a state of the system. 

The eigenstates of the free polariton Hamiltonian are plain
waves. The expectation value over the zero momentum state occupied
by many polaritons gives an infinite range coherence, $g^{(1)}\left(x\right)=1$.
In the next three sections we analyse each of the possible mechanisms separately to see how they restrict this behaviour.

\section{Finite momentum modes in a trap}

Here we obtain the single particle eigenstates of polaritons in the
harmonic trap. Then we evaluate $g^{(1)}\left(x\right)$ as the expectation value
with respect to a density matrix assuming a non-equilibrium distribution of the
polariton occupation numbers around the zero momentum state.

The single particle Hamiltonian for a polariton in a harmonic potential, 

\begin{equation}
H=\frac{p^{2}}{2m_{LP}}+U_{\textrm{harmonic}},\label{eq:H_hamonic}\end{equation}
is the one of the quantum harmonic oscillator. It is a well known
and solved model.

Following the standard approach to the diagonalisation problem we
search for a solution in form of a polynomial \cite{Schiff}. The
eigenfunction problem for the model Eq. (\ref{eq:H_hamonic}) leads
to Hermite's differential equation which is solved by 
\begin{equation}
\psi_{k}\left(x\right)=\frac{1}{\sqrt{2^{k}k!}}\left(\frac{2}{\pi}\right)^{\frac{1}{4}}\frac{1}{\sqrt{L}}e^{-x^{2}/L^{2}}H_{k}\left(\frac{\sqrt{2}x}{L}\right),\label{eq:psi_harmonic}\end{equation}
where $H_{k}\left(z\right)=\left(-1\right)^{k}e^{z^{2}}d^{k}e^{-z^{2}}/dz^{k}$
are the Hermite polynomials, $L=\sqrt{2\hbar/\left(m_{LP}\omega\right)}$
is the size of the system, and $k$ is the number of the energy level
that describes a finite momentum state of the 1D polariton. The complementary
eigenvalue problem can be solved via a Taylor expansion. As a result
the eigenenergies of the finite momentum states $k\neq0$ have a linear
spectrum
\begin{equation}
E_{k}=\omega\left(k+\frac{1}{2}\right).\label{eq:modes_spectrum}\end{equation}

A many polariton state, which is created in a pumping process, is
described by a density matrix. The expectation value in the correlation
function Eq. (\ref{eq:g1_def}) has to be evaluated as an ensemble
average, \begin{equation}
\left\langle \dots\right\rangle =\textrm{Tr}\hat{\rho}\dots,\label{eq:rho_average}\end{equation}
where $\hat{\rho}$ is a density matrix.

In a continuous wave experiment the signal is accumulated on the time
scale of seconds. The phase coherence of a polariton condensate is
limited by few hundreds of pico-seconds \cite{Love} which partitions
the long accumulation interval into a large ensemble of many very
short measurements. Thus the total signal is averaged over many realisations
with different distribution of the polariton occupation numbers and
the total number of polaritons which also changes slightly in time
due to the power fluctuation in the pumping laser \cite{Love}. As
the phase coherence is lost between different realisations the density
matrix, which describes such an ensemble, is diagonal in the representation
of the polariton occupation numbers, \begin{equation}
\hat{\rho}=\sum_{n_{0}n_{1}\dots}\rho_{n_{0}n_{1}\dots}\left|n_{0}n_{1}\dots\left\rangle \right\langle n_{0}n_{1}\dots\right|\end{equation}
Here $\left|n_{0}n_{1}\dots\right\rangle $ is a state with $n_{0}$
polaritons at the momentum $k=0$, $n_{1}$ polaritons at the momentum
$k=1$, etc and $\rho_{n_{0}n_{1}\dots}$ are the diagonal matrix
elements that describe a non-equilibrium distribution of the boson
occupation numbers.

Substituting Eqs. (\ref{eq:rho_average}) in Eq. (\ref{eq:g1_def})
and using the eigenfunction Eq. (\ref{eq:psi_harmonic}) we obtain
the first order coherence function as  sum over a set of  finite
momentum modes, \begin{equation}
g^{(1)}\left(x\right)=\sum_{k=0}^{\infty}\psi_{k}\left(x\right)\psi_{k}\left(-x\right)p_{k},\label{eq:g1_modes}\end{equation}
where $p_{k}=\sum_{n_{0}n_{1}\dots}n_{k}\rho_{n_{0}n_{1}\dots}/\sum_{n_{0}n_{1}\dots}\rho_{n_{0}n_{1}\dots}$
are weight factors, e. g. in thermal equilibrium they are given
by the Bose-Einstein distribution function. Note that $p_{k}$ are
not completely equivalent to the occupation number of the polaritons,
$n_{k}$, due to the normalisation of the correlation function in
Eq. (\ref{eq:g1_def}).

When only a few low momenta modes are excited with probabilities that are approximately  equal,
$p_{k}\simeq p_{l}$, the shape of the correlation function is almost
Gaussian with a reduced system size $\tilde{L}<L$, $g^{(1)}\left(x\right)\simeq e^{-x^{2}/\tilde{L}^{2}}$.
When many modes are excited with an arbitrary $p_{k}$ the shape is
arbitrary as polynomials in Eq. (\ref{eq:psi_harmonic}) of a high
order $k\gg1$ can approximate a wide range of different functions.

\section{Anderson Localisation}

Polaritons are subject to a random potential which originates from two sources.  One is disorder of the quantum well that influences the exciton component of polaritons. And another is  imperfections of the cavity mirrors that influences the photon component \cite{Kavokin2003,Keeling2004}. Here we consider a simplified model of the disorder characterised by a single correlation length Eq. (\ref{eq:disorder_def}) and we  assume that the wave length of the surface acoustic wave is smaller then the correlation length of disorder potential in the 2D heterostructure to model the strictly 1D system. If the latter is not true the polariton wire becomes a quasi-1D system which would give a longer localisation length. In this section we also neglect interplay between disorder and polariton-polariton interactions which can lead to a many body metal-insulator transition above a finite temperature \cite{Paul,Aleiner} and consider the low temperature case only.

Scaling analysis of the potential disorder problem \cite{scaling} shows that conductance
of a 1D system is zero at  low temperatures as all of the
zero momentum states become localised when the system size increases
to infinity. The lowest energy eigenstates of the single particle
model,

\begin{equation}
H=\frac{p^{2}}{2m_{LP}}+U_{\textrm{disorder}},\end{equation}
are localised with the exponential tails $\left|\psi\left(x\right)\right|\sim e^{\left(-\left|x\right|/\xi\right)}$,
for example see a review \cite{EversMirlin}. Therefore the first
order correlation function in Eq. (\ref{eq:g1_def}) evaluated with respect to
these states gives a finite coherence length with the same exponential
tail, \begin{equation}
g^{(1)}\left(x\right)\sim e^{-|x|/\xi},\label{eq:g1_disorder}\end{equation}
where, in general, the localisation length $\xi$ is not equal to
the correlation length of the random potential $l_{0}$. 

Here we do not perform a more detailed study of the features that are specific to the disorder in the polariton systems, e.g. \cite{Malpuech}, but only use the exponential tails of the correlation function as
a characteristic feature of the localisation mechanism.

\section{Polariton-Polariton interactions}

The Mermin-Wagner theorem \cite{MerminWagner} states that a long range
order in a 1D system is absent due to any finite-range exchange interaction which results in a large long-range quantum fluctuations. In this section we discuss the manifestation of this general statement in the specific system of interacting 1D polaritons.

The exciton component of a polariton has zero charge but a finite
dipole moment. Neglecting the photonic non-linearity, interaction between two polaritons has dipole-dipole nature thus it is short range. This interaction limits temporal coherence of the polariton condensate \cite{Love}. The spatial coherence, within the model of density-density interaction with the delta-function interaction, was was analysed in details in \cite{CarusottoCuiti,WoutersCarusotto}.


It was shown that the coherence length is finite 
which is manifested by the exponential tails of the coherence function,
\begin{equation}
g^{(1)}\left(x\right)\sim e^{-x/l_{c}},\label{eq:g1_interactions}\end{equation}
in accordance with the general Mermin-Wagner theorem \cite{MerminWagner}
that forbids an infinite range order in the 1D systems. It was also shown in \cite{WoutersCarusotto}
that the coherence length decreases with increase of the interaction
constant, $l_{c}\sim1/V_{0}$. For a typical GaAlAs microcavity the
estimate of $l_{c}$ was given as few hundred micrometers in an optimal
regime.

Note that the functional form of the  coherence function due
to the polariton interactions Eq. (\ref{eq:g1_interactions}) and due
to disorder Eq. (\ref{eq:g1_disorder}) coincides in the tails.

\section{Fitting of experimental data}

\begin{figure}
\centering\includegraphics[width=1\columnwidth]{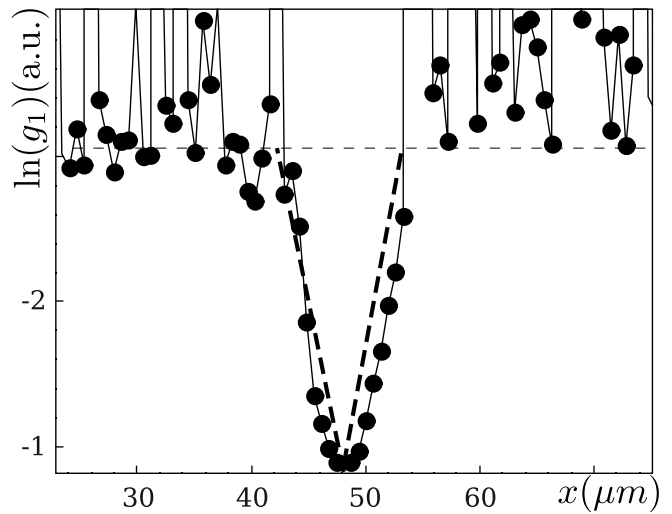}

\caption{\label{fig:g1_log}Coherence function $g^{(1)}$ measured along 1D wire
on the logarithmic scale. The thin dashed line marks the noise floor
and the thick dashed line is an exponential function.}
\end{figure}

\begin{figure}
\centering\includegraphics[width=1\columnwidth]{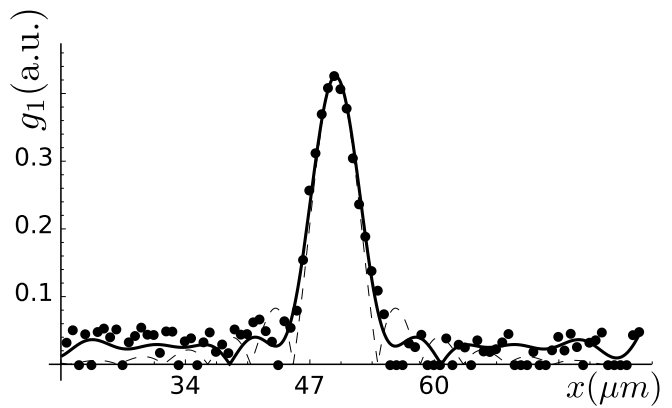}

\caption{\label{fig:modes_fit}Coherence function $g^{(1)}$ measured along 1D
wire on the normal scale. Thick line is the numerical fit using the
Eq. (\ref{fig:modes_fit}) with $p_{k}$ as the fitting parameters.
Thin dashed line is the plot of Eq. (\ref{fig:modes_fit}) with $p_{k}$
given by full triangles in Fig. \ref{fig:occupation_numbers}.}
\end{figure}
\begin{figure}
\centering\includegraphics[clip,width=1\columnwidth]{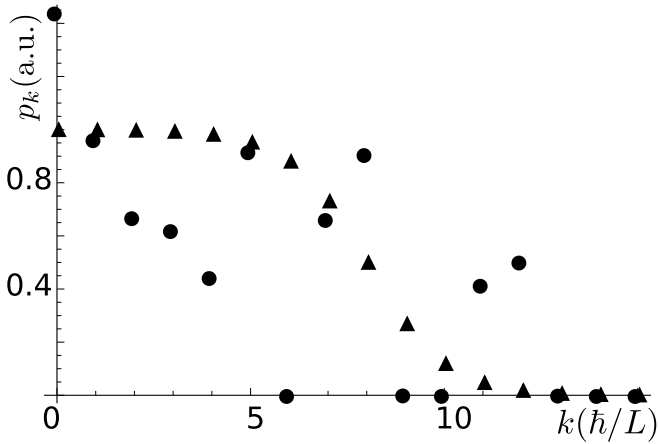}

\caption{\label{fig:occupation_numbers}Occupation numbers $p_{k}$ in Eq.
(\ref{eq:g1_modes}) to fit the experimentally measured $g^{(1)}$ :
full ellipsis - result of a numerical fit (think solid in Fig. \ref{fig:modes_fit}),
full triangle - a set of $p_{k}$ chosen by hand to occupy approximately
$10$ modes (thin dashed line in Fig. \ref{fig:modes_fit}).}
\end{figure}

Here we analyse experimental data \cite{Krizhanovskii,exp_data} on the coherence of polariton condensate along 1D wires.

In this experiment the 2D spot of the pumping laser defines the
system size and, therefore, limits the coherence length of the polaritons.
Then the microwave radiation is applied to confine the polaritons
to a wire-like geometry. It was observed that the coherence length in the direction
perpendicular to the wavefront was reduced from the system size to
the wave length of the surface acoustic wave but, unlike it was expected,
the coherence length in the unconstrained direction, parallel to the
wavefront, was also reduced.

We start from the qualitative analysis of the shape of the measured
coherence function, $g^{(1)}$, to identify the main mechanism that reduces the
coherence length along  the wires. The data on $g^{(1)}$ is presented in Fig.
\ref{fig:g1_log} on a logarithmic scale. The thick dashed lines marks an exponential function, which would correspond to the Anderson localisation or polariton-polariton interaction mechanisms from Sections 4 and 5, and serve as guides for the eye. Investigating both tails simultaneously, starting from the noise flow marked by the thin dashed line in Fig. \ref{fig:g1_log}, we find that the measured function has a super-linear character away from the maximum point \cite{symm}. Thus we conclude that the main mechanism is the non-equilibrium occupation of a set of the finite momentum modes from Section 3.

Having identified the mechanism we perform a more quantitative fitting.
We extract the 2D spot size from the data on $g^{(1)}$  which was measured without  the surface acoustic wave \cite{exp_data}. Gaussian fit gives the system size $L=16.3\;\mu m$.
Then we fit the data on  $g^{(1)}$, which was measured with the surface acoustic wave, by
the correlation function from  Eq. (\ref{eq:g1_modes})
using $p_{k}$ as the fitting parameters. The gradient descent method
gives the thick full line in Fig. \ref{fig:modes_fit}. The set of
$p_{k}$ for the thick line, which is presented by filled ellipses in Fig.
\ref{fig:occupation_numbers}, shows that approximately ten modes
are occupied.

This fitting of a single function with many parameters is not unique.
Another set of $p_{k}$ that is chosen ad hoc to occupy approximately the
same number of modes, triangles in Fig. \ref{fig:occupation_numbers},
also fits satisfactory the experimental data, thin dashed line in
Fig. \ref{fig:modes_fit}. It is not possible to extract the set of
$p_{k}$  from the data on $g^{(1)}$ uniquely but characteristic number of the
occupied modes is determinable.

We also analyse distribution of $p_{k}$ at thermal equilibrium. Bose-Einstein distribution, \begin{equation}
n_{k}=\frac{1}{e^{\epsilon_{k}\beta}-1}\end{equation}
where $\epsilon_{k}$ is the single particle spectrum of a harmonic
oscillator from Eq. (\ref{eq:modes_spectrum}) and $\beta$ is an
inverse temperature,  at low temperatures  gives the  distribution of the occupation numbers as $p_{k}\sim1/k$. It does not fit a satisfactory  the experimental
data. Thus we can conclude that polaritons are not in  thermal equilibrium.
This reflect the strong out-of-equilibrium nature of the constantly
pumped polariton system.

\section{Conclusions}

We have analysed different mechanisms that can limit the spatial coherence of a polaritonic condensate in 1D wires formed by the surface acoustic wave and have compared their predictions with the shape of the first order coherence function $g^{(1)}\left(x\right)$ measured in \cite{Krizhanovskii}. From a qualitative analysis of the experimental data we have found that the main mechanism is the non-equilibirum occupation of a set of the finite momentum modes. Out of the three effects that we considered: Anderson localisation
and the polariton-polariton interaction give an exponential tails of $g^{(1)}\left(x\right)$, the finite  momentum
modes  occupation  gives a Gaussian shape. In the experimental data \cite{exp_data}   the shape is Gaussian, up to the noise floor. Performing a more quantitative fit,
using the occupation number as the fitting parameters, we have found that
the number of the excited modes in the experiment \cite{Krizhanovskii}
is $\sim10$. 

From the analysis in the present paper it is advised to use a static method of the polariton confinement to extend the spatial coherence in the 1D geometry. The finite momentum modes are most probably excited due to  dynamical nature of the acoustic lattice that interacts with the zero momentum polaritons, populated indirectly in  the OPO setup, and scatters them to finite momenta. Therefore, for example, etching the whole planar structure to form a wire will remove  the main source of dephasing that limits currently the coherence in the spatial domain. Such structures were recently produced \cite{Wertz} and a coherent  propagation over a long distance  of tenth of micrometers in these structures was reported \cite{Wertz}.   

Acoustic lattice induced regime of a 1D condensate, which was identified in Section 6, opens a new way to study distinct non-equilibirum properties of the polaritons. A better insight can be obtain in more complex frameworks,  such as \cite{eastham2001,carusotto2007,savona2008}, that can capture finer details and can provide a further understanding of the non-equilibirum processes in microcavities.

\section{Acknowledgements}

We thank M. S. Skolnick and D. N. Krizhanovskii for discussions and
for their experimental data on the spatially resolved $g^{(1)}$  function which was communicated
to us. This work was supported by the EPSRC Programme Grant EP/G001642/1.


\begin{thebibliography}{[1]}
\bibitem{polariton_phonon} We do not consider the effect of polariton-phonon interactions on the polariton coherence as this interaction is slow $\sim 100$ ps \cite{Tassone} compared to the system life-time $\sim 5$ ps \cite{Balili}.
\bibitem{symm} Note that noticeable skewness in the measured  $g^{(1)}(x)$ in Figs. 2 and 3 is a spurious artefact of the measurement. The conclusion about the super-exponential character of the tails was drawn after a symmetrisation.
\end{thebibliography}
\end{document}